# Non-linear Instabilities

Lun-Shin Yao
Arizona State University
Tempe, AZ 85287

**Abstract**

Wave resonance is the fundamental mechanism of non-linear instabilities of fluid flows, and affects the long-time evolution of fluid motions and other physical problems described by non-linear differential equations. Some significant consequences, not well known and completely different from linear instabilities, are summarized in this paper from the studies of past two decades.

## 1. Introduction.

Substantial progress of fluid mechanics has been made in the past century. Theories of linear instability and weakly non-linear instability [Landau 1959, Stuart 1960, and Watson 1960, Philips 1960] are landmark contributions in fluid mechanics. The fact, that the progresses of fluid mechanics are made by linearized analyses and small-scale experiments has been repeatedly mentioned, but often been overlooked. Modern computers provide convenient tools to solve the Navier-Stokes equations directly without severe simplification, which is required by linearized analyses. Computation becomes a routine exercise in solving practical problems in fluid mechanics, in estimating linear-instability limits and finite-amplitudes of weakly instabilities. Better computing devices do not, however, pull equal progress to advance essential concepts in fluid mechanics. The importance of non-linearity of the Navier-Stokes equations and its consequence were covered by the dust of concepts learned from *linearized* theories in fluid mechanics. This may due to the fact that we are lacking of capability to solve non-linear differential equations *analytically* in order to find out the true characters of non-linearity and the linked physics.

It is well known that many difficult phenomena associated with the Navier-Stokes equations are subtle and subject to different interpretations. In this regard, four issues are of interest in numerically solving the Navier-Stokes equations. The first is a linear-stability analysis; the next two are related to nonlinear interactions (instabilities); and the last is concerned with the lack of convergent numerical chaotic solutions and error evolutions. The understanding of these issues is not straightforward since no general analytical solutions exist. The purpose of this paper is to summarize the results of some recent progress on these issues made in the past two decades and other related matters, which are not well known.

The spectral method used is described in section 2. In section 3, important consequences of non-linear instability, which are rather different from those from a linearized study, are summarized. The issues involve the behaviors of solutions when the value of the governing parameter is above the critical state; in this case, nonlinear interactions among disturbances of non-zero amplitudes become important. Such behaviors are the key factors in the understanding of various transition processes in fluid flows. Classical treatments of nonlinear interactions can be traced to the pioneering efforts of Landau [1959], Stuart [1960], and Watson [1960] for shear flows, and Philips [1960] for water waves; unfortunately, their studies are focused mainly in a single wave, or single wave packet. In section 3, it is shown that a linearly stable or unstable initial condition always excites the entire set of harmonics appearing *simultaneously* in its Fourier components, no matter how small their amplitudes. Consequently, the solution is nonlinearly stable at the critical state estimated by a linear theory. This is because that the influence of the linear term vanishes for the critical wave, which can transfer energy simultaneously to its more dissipative harmonics. Another normal consequence is that a *single-wave* solution does not exist for a non-linear partial differential equation since the entire set of harmonics is excited. This result provides a simple explanation as to why a long wave can trigger many short waves simultaneously, and why kinetic energy is not necessarily transferred from long waves to short waves in cascade.

We have previously demonstrated [Yao 1999] that the nonlinear terms of the Navier-Stokes equations can be interpreted as forced and resonant vibrations, and can induce "kinetic energy" transfer among waves. Here energy is expressed by the square of the wave amplitude. Since forced transfers do not cause significant transfer of energy among individual waves, they do not change the spatial structure of the solution. However, they do excite all wave harmonics, thereby providing a starting point for resonant energy transfers. In contrast, the energy transferred due to resonance is substantial, and can lead to observable changes in the solution structure. Consequently, the energy transfer associated with the nonlinear terms cannot be ignored as long as wave amplitudes are not zero. One numerical example is presented in section 3 to illustrate the existence of such behaviors for the Navier-Stokes equations. These numerical results also show that the energy can be transferred to sub-harmonics. It also shows that the long-time multiple solutions at a given Reynolds number are sensitive to initial conditions, which suggests that the classical principle of "dynamic similarity" is strictly valid only for laminar flows. This relates to the first of two most misunderstood facts among many issues about non-linear differential equations. It is generally viewed that a *sensitivity-to-initial* condition is a *necessary* and *sufficient* condition for the existence of chaos. However, this property is also noted in the solutions of all non-linear differential equations when the value of their governing parameters is larger than their critical value. Consequently, it cannot be argued that this sensitivity is a sufficient condition for chaos. One can also relay that any turbulent-flow problem may have many sets of statistic structures depending on environment and initial conditions.

In section 4, the non-convergence of numerical computations for unstable flows is briefly described. The amplification of numerical error for non-linear differential equations is determined by non-linear instabilities, not linear instabilities. The second most disregarded fact about non-linear differential equations is: *no computed chaotic or turbulence solutions, which are independent of the integration time-step; consequently all are numerical errors* [Teixeira, Reynolds and Judd 2007, Yao and Hughes 2008, Yao 2010b]. The reality is that numerical solutions of chaos or turbulence do not exist and are beyond our current mathematics; even though, those phenomena have indeed been observed in our lives.

**2**. **Nonlinear Theory (Fourier-Eigenfunction Spectral Formulation)**

The following is a summary of the formulation by Yao [1999, 2007, 2009], and Yao and Ghosh Moulic [1994, 1995a, 1995b], and Ghosh Moulic and Yao [1996]. A fully-developed mean flow in cylindrical polar coordinates is used as an example to illustrate the nonlinear theory. This limitation of fully-developed mean flows can be removed with a substantial increase in the computational time. The formulation is for a temporal flow development, but can be easily converted to a spatial flow development. Various weakly non-linear theories for single wave can be interpreted as a simplified version of our general theory [Yao 2007, 2009].

The velocity components, (u, v, w) and temperature, θ, are decomposed into mean values plus disturbances. The magnitude of the disturbances is *not* necessarily smaller than that of the mean values. They are

$$\mathbf{u} = (u, v, w, \theta) = (\mathbf{U}(\mathbf{r}) + \mathbf{u}', \Theta_0(r) + \theta') = (u', v', W_0(r) + w', \Theta_0(r) + \theta'), \quad (1)$$

where $W_0$ and $\Theta_0$ are the mean axial velocity and temperature. The disturbance is expressed as

$$\mathbf{u}(\mathbf{r}, t) = \mathbf{U}(\mathbf{r}) + \int_{-\infty}^{\infty} \sum_{n=-\infty}^{\infty} \sum_{m=1}^{\infty} A_m(k, n, t) \tilde{\mathbf{u}}_m(k, n, r) e^{i(kz + n\phi)} dk, \quad (2)$$

where $A_m$ is the amplitude density function, and $\tilde{\mathbf{u}}_m$'s are the linear stability eigenfunctions. The terms in (2) corresponding to $n = k = 0$ represent the modification of the mean flow and temperature due to the evolution of disturbances. This implies that the mean flow can also be treated as one wave with infinite-long wave length and zero frequency. Such a feed back is missing from all weakly non-linear theories for a single wave, or for a narrow wave band. Furthermore, all waves are formulated in such a way they are represented fully and are not as a group. This is an important property, which allows them to non-linear interact among themselves. This complete representation of velocities ensures the conservation of kinetic energy transferred among all waves, leads to proper evolution of non-linear instabilities, and results multiple solutions for the Navier-Stokes equations. The selection of final solution depends on the initial conditions, which are unavoidably influence by environment. The continuous spectra can also be included in this formulation. Substituting (2) into the Navier-Stokes equations and projecting along the direction of the adjoint eigenfunctions result in

$$\frac{\partial A_m}{\partial t} + i\omega_m A_m = \sum_{m_1=1}^{\infty} \sum_{m_2=1}^{\infty} \sum_{n_1=-\infty}^{\infty} I(k, n, m, m_1, m_2, n_1, t), \quad (3)$$

where

$$I = \int_{-\infty}^{\infty} b(k_1, k - k_1, n_1, n - n_1, m_1, m_2, m) A_{m_1}(k_1, n_1, t) A_{m_2}(k - k_1, n - n_1, t) dk_1,$$

and the b's are constants of wave interactions. The linear terms of the Navier-Stokes equations represent the mean-flow convection of the disturbances, the distortion of the disturbances by the mean-flow stresses and the body forces, and diffusion. In the generalized coordinates of eigenfunctions, they are reduced to a single term, which determines the growth or decay of the wave due to linear energy transfer. This is a unique advantage of Fourier-Eigenfunction Spectral formulation compared to other spectral methods. The equation (3) is divergent free for incompressible fluid; the difficult associated with the decoupling of pressure in the continuity equation does not exist in the current formulation. The consequence is a substantial saving of CPU time.

For a fixed wavenumber, $\omega_m^i$ is more negative for larger m. This indicates that eigenfunctions for larger m have more dissipation, or it can be interrelated as a smaller eddy. But, this eddy is not in a spherical shape and is elongated. This introduces anisotropy into the flow structure. For laminar flows, the feedback of small-scale motions to large-scale structures is negligible. For instability flows, the magnitude of small-scale motions may not be much smaller than that of large-scale structures. Their structures are determined by the stability characteristics of the mean flow. This fact has been used to correlate many experimental data.

The eigenfunction expansion (2) has reduced the continuity, momentum, and energy equations to the system of integro-differential equations for the amplitude density functions without any approximations. Thus, the solution of equations (3) represents an exact solution of the Navier-Stokes equations. It is worth pointing out again that one of the major difficulties in the numerical solution of the incompressible Navier-Stokes system is the simultaneous enforcement of the no-slip boundary conditions and the incompressibility constraint. Since the base functions used in the expansion (2) are solutions of the linearized Navier-Stokes equations, they individually satisfy the incompressibility constraint as well as the boundary conditions. This is equivalent to project the computational domain to a sub-domain where the expansion (2) automatically satisfies the boundary conditions and the continuity equation. Thus, the numerical solution of the system of equations (3) is much simpler than the numerical solution of the Navier-Stokes equations. It may also be noted that straightforward evaluation of the convolution product representing the nonlinear terms in equation (3) is inefficient if the number of terms in the truncated eigenfunction expansion used in the numerical solution is large. However, pseudo-spectral evaluation of the convolution product can make the numerical solution of the equations (3) a viable efficient alternative to the numerical solution of the Navier-Stokes equations. A preliminary study shows that the required

CPU time can be as little as one sixth of that needed to solve the Navier-Stokes equations by a Fourier-Chebyshev collocation spectral method [Yao and Ghosh Moulic 1994]. The weakness of the method is that the eigenfunctions are for a selected problem and Re only.

The non-linear terms in (3) represent the forced and resonant transfer of kinetic energy among all waves. The forced transfer can produce new waves, but provides minimum energy transfer. This mechanism is important for creating sufficient waves to allow harmonic energy transfer. Substantial kinetic energy transfer can happen when the resonance conditions meet. This resonance mechanism was first identified by Phillips [1960] for surface waves, and for surface and internal waves. Our analysis indicates that the velocity field is represented as (2), the principle of resonant transfer is pertinent to whole flow field. Our study generalizes the Phillips theory, and provides a simple explanation why surface waves and internal waves can resonantly exchange energy.

The resonance conditions have been summarized by Yao [1999] and reproduced below.

1. The non-linear terms in (3) include resonance interaction of three, four, and many waves [see also, Yao and Ghosh Moulin, 1995a, 1995b].

2. The resonance condition requires that the frequencies of waves satisfy the exact relation of the wave numbers. If the waves are phase-locked, then the resonance condition is automatically satisfied. Two commonly observed cases are harmonic and sub-harmonic resonances. The harmonic resonance of three waves occurs when a wave interacts with itself and transfers energy to its first harmonic. Let us represent this symbolically as $2\mathbf{k} = \mathbf{k} + \mathbf{k}$, where $\mathbf{k}$ is the vector of wave number. The left-hand side of the equality sign is the new wave and the right-hand side is the interaction of existing waves. The sub-harmonic resonance is $\mathbf{k} = 2\mathbf{k} - \mathbf{k}$. This shows that the sub-harmonic wave must exist before this resonance can appear, since $\mathbf{k}$ shows up on the right-hand side of the equal sign. The numerical results of Yao and Ghosh Moulin (1994) confirmed it. In reality, noise waves of all wave numbers can exist, but their amplitudes may be too low to be noticeable. The harmonic and sub-harmonic resonances have been observed in turbulent mixing layers (Dimotakis and Brown 1976, Browand and Troutt 1980, Oster and Wygnanski 1982, Huang and Ho 1990), wakes and jets (Crow and Champagne 1971, Moore 1976).

3. The earliest recognized resonance group in fluid dynamics was $\mathbf{k} = \mathbf{k} + \mathbf{k} - \mathbf{k}$. This is the self-interaction of resonant quartets that transfer energy from waves to the mean flow to balance the linear growth of waves. This resonance condition is always satisfied. This is the resonance considered by most weakly non-linear instability theories (Stuart 1960, Watson 1960, Stewartson and Stuart 1971). The modification of the mean flow is not considered in those weakly non-linear theories, since they are formulated for a single wave. This is why multiple solutions have not detected by the early instability study.

4. All waves can always interact *non-linearly* with the mean flow. Symbolically, this is **k** = **k** + **0** which forms a resonant trio. This is the most important one for transient flows and linear instability is a special case for this kind of resonance. The existence of this resonance allows all waves interchange energy indirectly via their interaction with the mean flow.

5. With limited computational results, it is not possible to conclude with confidence that in-commensurate waves, which do not satisfy resonance conditions, cannot co-exist in a flow. But, both can exchange kinetic energies with the mean flow resonantly. A typical example of such a case is that two solitons, two groups of in-commensurate waves, pass each other without changing their shapes, or exchanging kinetic energy between them.

**3. Consequences of Non-Linear Instabilities:**

Since there are many waves can exist in a flow, some previous fluid mechanic principles, established by classic single-wave theories, have rather limited scope and require modification.

1. The linear instability theory is the solution of (3) by ignoring the non-linear interaction terms [Yao 1999, 2007]. The instability of each wave is decoupled from other waves, and can be studied independently. This simplification drastically reduce computational effort in evaluating the critical Reynolds number, $Re_{cri}$. The decoupling in the linear instability computations neglect a small amount energy, which is transferred to other more stable waves. Consequently, the flow is stable at the critical Reynolds number, estimated by the linear-stability analyses. The true critical Reynolds number should be slightly above that estimated by the linear-stability analyses.

2. Inside the unstable region, near the linear-stability boundary in the space of Re and wave number, the waves are stable. This is the well known Benjamin-Feir and Eckhaus side-band instability [Eckhaus 1965, Benjamin and Feir 1967, Stuart & DiPrima 1978]. They applied the weakly-instability analysis by perturbing two side-band waves with small difference in the wave number. Such a small perturbation is required by applying a weakly-instability analysis. In our computation, we found the same phenomena without presenting any side-band waves. It is simply due to the fact that waves in this narrow band, near the linear-stability boundary, do not receive sufficient energy from the mean flow, but simultaneously transfer more energy to their harmonics, or sub-harmonics. This results their amplitudes declining. The outcome is their harmonics, or sub-harmonics, which seat in the middle of the unstable region, become the dominant wave.

3. From the above description, it is easy to conclude that no stable flow is possible for $Re > Re_{cri}$. This is contrary to the conclusion from a linear-stability analysis that, for $Re > Re_{cri}$, only when a wave number is within the unstable range, the flow is unstable. This situation can be better demonstrated by a numerical example.

Results are for an infinite long Taylor rotational column with radius ratios $h = 0.5$, which corresponds to the apparatus used in the experiments of Snyder (1969). Linear stability analysis indicates that for $Re = 88.1$, circular Couette flow is unstable to rotationally symmetric disturbances with axial wavenumbers lying between 1.6 and 5.6. The system of integro-differential equation (3) was solved numerically with different initial conditions using 20 eigenmodes for each wavenumber, which is, using 20 terms in the eigenfunction expansion given by equation (2). The integrals in (3) were discretized by the trapezoidal rule using a uniform mesh size $\Delta k = 0.25$. The infinite range of integration was truncated to $-12 \leq k \leq 12$, which was found to be adequate as the kinetic energies of the high wavenumber modes were negligible. At $Re = 88.1$ and $h = 0.5$, circular Couette flow is linearly stable to non-rotationally symmetric disturbances, and the computations for this case were done assuming the disturbance to be rotationally symmetric in order to save computer time

Figure 1 shows the results of a numerical solution of the integro-differential equations (3) for $h = 0.5$ and $Re = 88.1$ in which the initial disturbance consists of a single dominant mode with wavenumber $k = 3$. The evolution of the kinetic energies of the dominant wave components is plotted in Figure 1. For rotationally symmetric flows, the kinetic energy of the Fourier component with axial wavenumber k is given by

$$E(k,t) = \begin{cases} \int_{r_i}^{r_o} r\left[|\hat{u}|^2 + |\hat{v}|^2 + |\hat{w}|^2\right] dr, & k \neq 0 \\ \dfrac{1}{2} \int_{r_i}^{r_o} r\left[|V + \hat{v}|^2 - V^2\right] dr, & k = 0 \end{cases}. \tag{4}$$

Equation (4) accounts for the energy in both modes ±k. The kinetic energy of the circular Couette flow is subtracted from the mean-flow kinetic energy in so that E (0, t) represents the kinetic energy associated with the distortion of the mean flow. The mode k = 3 is linearly unstable, and grows initially at the rate predicted by linear stability theory. Nonlinear interactions generate harmonics of the wave k = 3 and induce a mean-flow distortion (k = 0). As the amplitude of the mode k = 3 increases, nonlinear effects become important and alter the growth rate, causing the mode to decay and eventually reach an equilibrium state. The energy is drawn out from the mean flow to support the disturbance wave and its harmonics. Table 1 shows the amplitudes $\overline{A}_m(k) = \Delta k \cdot |A_m(k, n = 0)|$ of the first ten eigenmodes of the different harmonic components of the disturbance in this equilibrium state, where $|A_m(k,n)|$ denotes the magnitude of the complex amplitude density function. Table 2 gives the corresponding frequencies $\dfrac{d\beta_m(k)}{dt}$ of the various eigenmodes, where $\beta_m = \tan^{-1}\left(A_m^I / A_m^R\right)$ is the phase angle of complex amplitude-density function, $A_m(k, n)$. A glance at Table 2 reveals that most of the eigenmodes have zero frequencies, that is, they represent a stationary disturbance. However, some of the eigenmodes have non-zero frequencies. These eigenmodes represent a time-periodic disturbance. A close examination of Table 2 shows that the time-periodic eigenmodes for a given wavenumber occur in pairs which have frequencies of the same magnitude but opposite sign. For instance, the 4th and 5th eigenmodes of the wavenumber k = 3 have frequencies of 0.0105 and -0.0105 respectively. These two eigenmodes represent traveling waves which move in opposite directions with the same wave speed. Other such pairs are formed by the 4th and 5th eigenmodes of the wavenumber k = 6, the 6th and 7th eigenmodes of the wavenumbers k = 6, 9 and 12, and the 8th and 9th eigenmodes of the wavenumber k = 12. These eigenmodes highlighted in Table 2 can carry disturbances from the ends of an annulus into its interior. Table 1 indicates that the eigenmodes of each of these pairs have the same amplitude. For instance, the 4th and 5th eigenmodes of the wavenumber k = 3 have the same amplitude 0.004497, the 4th and 5th eigenmodes of the wavenumber k = 6 have the same amplitude 0.00512 etc. The superposition of all the eigenmodes results in a standing wave pattern. Any disturbance which destroys the symmetry of the standing waves can result in the addition of components of spiral flows. Table 1 show that the amplitudes of the time-periodic eigenmodes are an order of magnitude smaller than the amplitude of the stationary Taylor vortex cell. This may explain why spiral flows have not been identified in experiments on rotationally symmetric Taylor-vortex flows due to their small amplitudes.

Figure 2 shows the results of a numerical solution of the integro-differential equations (3) at Re = 88.1 and h = 0.5 starting with a single dominant mode with wavenumber k = 1.75. The mode k = 1.75 is linearly unstable. However, unlike the previous case, the mode k = 1.75 decays to zero instead of growing to a finite-amplitude equilibrium state, while its harmonic k = 3.5, excited through nonlinear wave-interaction, grows and reaches a supercritical equilibrium state. This result is in agreement with the Eckhaus and Benjamin-Feir side-band instability, but different from the reason provided by the weakly non-linear theory. The equilibrium state in this case also is a standing-wave pattern with a dominant stationary Taylor vortex cell and some oscillatory modes of much smaller amplitudes. However, the dominant wave in this case has a wavenumber k = 3.5 in contrast to the previous case which had a dominant wavenumber k = 3.

The results of Figures 1 and 2 indicate that the equilibrium state of the flow is not uniquely determined by the Reynolds number and radius ratio of the cylinders, but depends on the waveform of the initial disturbance. Multiple stable equilibrium states can occur at the same Reynolds number. We have computed several stable rotationally symmetric Taylor-vortex flows for h = 0.5 and Re = 88.1 by integrating (3) starting with initial conditions consisting of a single linearly unstable mode of small initial amplitude. In all of these cases, the final equilibrium state was a monochromatic standing wave with a single dominant wavenumber and its harmonies. The standing wave pattern consists of a dominant stationary Taylor-vortex cell and some low- frequency oscillatory modes of much smaller amplitudes. Our computations indicate that the range of wavenumbers for stable supercritical Taylor vortices is narrower than the span of the neutral curve of linear stability theory. When the initial disturbance consists of a single dominant mode with wavenumber inside this narrow band, the initial wave remained the dominant wave in the equilibrium state, as in the case illustrated by Figure 1. When the initial disturbance consists of a single dominant mode with wavenumber outside this narrow band but within the unstable region of linear theory, the initial wave decayed, the energy being transferred to a wave inside the narrow band, which is excited through nonlinear wave-interaction, as in Figure 2. This indicates that the phenomenon, mentioned above, usually associated with sideband instability, is not necessary due to side-band instability. This is a typical example to show that using a *weakly* non-linear analysis to explore a phenomenon associated with non-linear differential equations can be misleading.

We have examined the stability of some of these equilibrium states to finite-amplitude disturbances. Figures 3 and 4 show the results obtained by perturbing a stable supercritical Taylor-vortex flow in which the critical wavenumber k = 3.25 was the dominant wave with a disturbance which consists of a linearly unstable eigenmode with wavenumber k = 3.5. Figure 3 presents the results of a computation in which the initial perturbation was given amplitude of 0.1. The figure indicates that the wave k = 3.25 decays to zero, while the wave k = 3.5 becomes the dominant wave in the equilibrium state. Figure 4 presents the results of a similar computation in which the initial perturbation was given amplitude of 0.075. In this case, the perturbation decays to zero, and the critical wave remains dominant in the equilibrium state. Figures 3 and 4 indicate that the dominant wavenumber in the equilibrium state can be shifted by a disturbance of sufficiently large amplitude. This phenomenon can have practical significance in engineering applications. It is easy to conclude that multiple solutions are a normal property for non-linear differential equations [Yao 2009].

Figure 5 shows the projection of the velocity vectors for the equilibrium state depicted by Figure 2 onto the r-z plane. The first and second harmonic components of the velocity vectors are also shown in Figure 3, along with the total velocity vectors. The figure indicates that the velocity vectors of the second harmonic near the inflow boundary of the Taylor-vortex are opposite in direction to the velocity vectors of the first harmonic. The velocity vectors of the second harmonic near the outflow boundary of the Taylor-vortex, on the other hand, are in the same direction as that of the first harmonic. Thus, the second harmonic component of the disturbance, generated through non-linear wave-interaction, tends to reduce the radial velocity of the inflow boundary jet between the Taylor vortices and increase the radial velocity of the outflow boundary jet. Thus, the outflow boundary has a larger radial velocity than the inflow boundary.

It is worth noting that the results obtained using 15 and 20 eigenfunctions completely agree within the accuracy of our computation. A convergent result by a Fourier-Chebyshev spectral method, comparable to the expansion of 15 or 20 eigenfunctions, requires a minimum of 33 Chebyshev polynomials. The results presented in Figures 2 and 3 are obtained using 20 eigenfunctions in the expansion (2). The required number of eigenfunctions in the expansion (2) for different Reynolds numbers can be determined by demanding that the amplitudes of highest few eigenmodes are sufficiently close to zero. This means that the expansion (2) must be convergent in order to provide physically meaningful results.

We also found that, if the expansion (2) is severely truncated to, say, five or fewer eigenfunctions, the results are qualitatively different from the convergent results for the same initial conditions; sometimes, the results may become numerically unstable and diverge. This is due to lack of dissipative mechanism in a severely truncated system. The higher eigenmodes corresponding to larger $\omega_m^I$ are more dissipative in nature, and a sufficient number of eigenmodes is needed for *numerical stability*. Physically, this implies that a substantial dissipation occurs non-isotropically within the range of wavenumbers for energy-containing eddies. Geometrically, one visualizes that a substantial dissipation occurs in thin layers whose sizes along the axial and the azimuthal directions are comparable to the energy-containing eddies. The equilibrium waves of larger wavenumbers are the harmonics of the dominant wave, are dynamically inert and are not particularly more dissipative. Their small amplitudes are the consequence that little energy is transferred to them non-linearly. Similar conclusion can be made for the sub-harmonics of the dominant wave.

More computational results for Taylor instability and mixed convection in an annulus can be found from the above referenced papers by Yao and his coworkers.

## 4. Evolution of Numerical Errors for Non-linear Differential Equations

Von Neumann established that discretized algebraic equations must be *consistent* with the differential equations, and must be *stable* in order to obtain *convergent* numerical solutions for the given differential equations [Yao & Hughes 2008]. Flows, above the critical Reynolds number are unstable. It is not possible to construct a stable discrete algebraic equation. We found, from our experience, that instability flow can indeed be computed, if the Reynolds number is not much larger than its critical value. Since the available computer time is very limited, we did not find out the upper limit of Reynolds numbers that no convergent numerical solutions are possible. We do know such a limit is far below the *transition* Reynolds number.

It is worthy to point out that the principle of resonant transfer of kinetic energy is very similar between the Navier-Stokes equations, the one-dimensional Kuramoto-Sivashinsky equation, and the Lorenz equations [Yao 2010a]. From our study of a simple model non-linear differential equation, one-dimensional, Kuramoto-Sivashinsky equation [Yao 1999, 2007], we realize that numerical solutions divergent rather fast when the value of the parameter increases above its critical value. We have repeatedly tried to solve the model equation numerically, and found, three-year after the papers were published, that the chaotic solutions presented in the appendix of [Yao 1999, 2007] are numerical errors, since they are integration time-step dependent.

Traditionally, we state that numerical errors are amplified *exponentially* for unstable flows. This is the conclusion from linear instability analyses of non-linear differential equations. In this paper, we have demonstrated a small perturbation can be amplified much more complex for non-linear differential equations than their counterpart for linear differential equations. Not only errors can be amplified; their harmonics and sub-harmonics can also be generated. Sometimes, the amplitude of harmonics and sub-harmonics can be much larger than those of numerical errors themselves. A small error can pull computational results close to a different possible solution from the one that was originally sought. This is a problem area, which has never been seriously studied [Teixeira, Reynolds and Judd 2007, Yao and Hughes 2008, Yao 2010b], since our knowledge about non-linear instability is still in its toddler stage.

## 5.  Conclusion

I hope I have clearly explained why and how non-uniqueness is a generic property for all fluid flows. Reynolds number alone is insufficient to uniquely determine a flow field and its transport properties. Low-amplitude environmental perturbations can have profound effects on the determination of the equilibrium state. For problems near the onset of instability, the required modification of the dynamic similarity ensured by the Reynolds number might be small. But for a fixed Reynolds number practically above its critical value, the variation of engineering data, such as flow resistance and Nusselt numbers, can be substantial. In spite of the fact, it is clear that the involved new physics of multiple solutions is wave resonance; unfortunately, possible ranges of flow variations and problem dependent, are unknown.

Finally, note that the accurate numerical computation of unstable flows, such as a flow with the Reynolds number much larger than its critical value, is not possible with any discrete numerical methods: Any such computational results are incorrect and simply the accumulation of truncation errors [Yao 2007, Yao 2008, Yao and Hughes 2008a, 2008b]. This result rules out the possibility of using current numerical methods to study flows at Reynolds numbers much larger than their critical values. Unfortunately, this includes the major parameter range in which multiple solutions exist. This also reveals one of the primary reasons why the study of multiple solutions and turbulence is so difficult. It is, however, clear in concept that non-linear instability analyses have much larger scope than that of linear instability analyses. Their differences are by no means small. Only non-linear instability theory can provide much deeper and more thorough and accurate understanding of fluid mechanics and other physics associated with non-linear differential equations.

Table 1.  Amplitudes of the different eigenmodes for Re = 88.1 and h = 0.5 for a single initial mode with k = 3.

| m | k = 0 | k = 3 | k = 6 | k = 9 | k = 12 |
|---|---|---|---|---|---|
| 1 | 0.3641E-01 | 0.8650E-01 | 0.1772E-01 | 0.3229E-02 | 0.8255E-03 |
| 2 | 0.1957E-11 | 0.8715E-02 | 0.1145E-02 | 0.1072E-03 | 0.1986E-03 |
| 3 | 0.4582E-01 | 0.4474E-02 | 0.5770E-02 | 0.8343E-03 | 0.3437E-04 |
| 4 | 0.3007E-11 | **0.4497E-02** | **0.5120E-02** | 0.5860E-03 | 0.5571E-04 |
| 5 | 0.1273E-02 | **0.4497E-02** | **0.5120E-02** | 0.4384E-03 | 0.9338E-04 |
| 6 | 0.1076E-11 | 0.5323E-03 | **0.1131E-02** | **0.2789E-03** | **0.7535E-04** |
| 7 | 0.1416E-02 | 0.5030E-04 | **0.1131E-02** | **0.2789E-03** | **0.7535E-04** |
| 8 | 0.9047E-12 | 0.2625E-03 | 0.5979E-03 | 0.3127E-03 | **0.6523E-04** |
| 9 | 0.8591E-04 | 0.2808E-03 | 0.1584E-03 | 0.1638E-03 | **0.6523E-04** |
| 10 | 0.6120E-13 | 0.2142E-04 | 0.5842E-04 | 0.4279E-04 | 0.1827E-04 |

Table 2.  Frequencies of the different eigenmodes for Re = 88.1 and h = 0.5 for a single initial mode with k = 3.

| m | k = 0 | k = 3 | k = 6 | k = 9 | k = 12 |
|---|---|---|---|---|---|
| 1 | 2.45E-16 | 3.44E-12 | 6.89E-12 | 1.03E-11 | 1.38E-11 |
| 2 | 2.45E-16 | 3.44E-12 | 6.89E-12 | 1.03E-11 | 1.38E-11 |
| 3 | 2.45E-16 | 3.44E-12 | 6.89E-12 | 1.03E-11 | 1.38E-11 |
| 4 | 2.45E-16 | **0.0105** | **-0.0097** | 1.03E-11 | 1.38E-11 |
| 5 | 2.45E-16 | **-0.0105** | **0.0097** | 1.03E-11 | 1.38E-11 |
| 6 | 2.45E-16 | 3.44E-12 | **-0.0123** | **-0.0214** | **0.0235** |
| 7 | 2.45E-16 | 3.45E-12 | **0.0123** | **0.0214** | **-0.0235** |
| 8 | 2.45E-16 | 3.44E-12 | 6.89E-12 | 1.03E-11 | **0.0151** |
| 9 | 2.45E-16 | 3.44E-12 | 6.89E-12 | 1.03E-11 | **-0.0151** |
| 10 | 2.45E-16 | 3.44E-12 | 6.89E-12 | 1.03E-11 | 1.38E-11 |

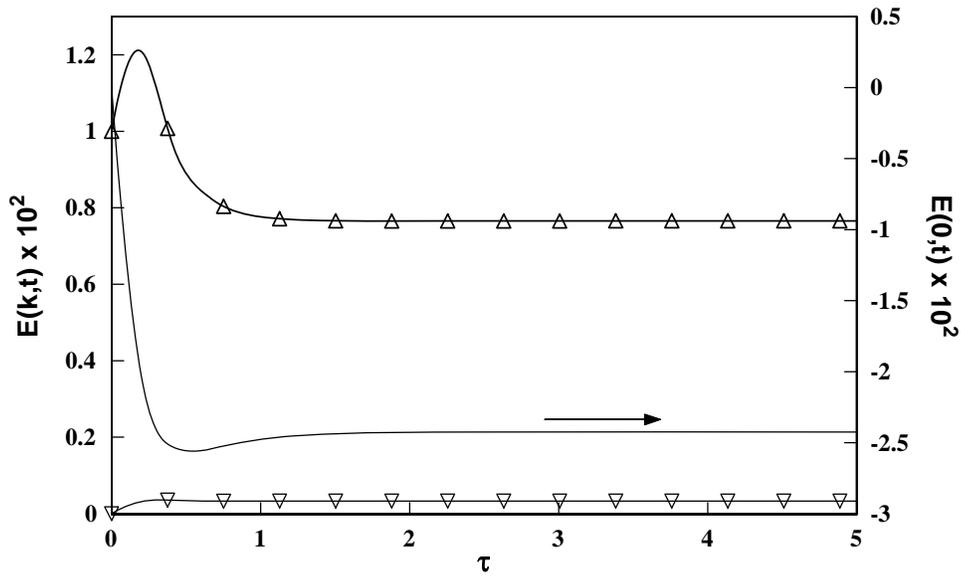

Figure 1. Evolution of the kinetic energy for Re=88.1 and h=0.5 for a single initial mode k = 3.

—— k = 0   —△— k = 3   —▽— k = 6

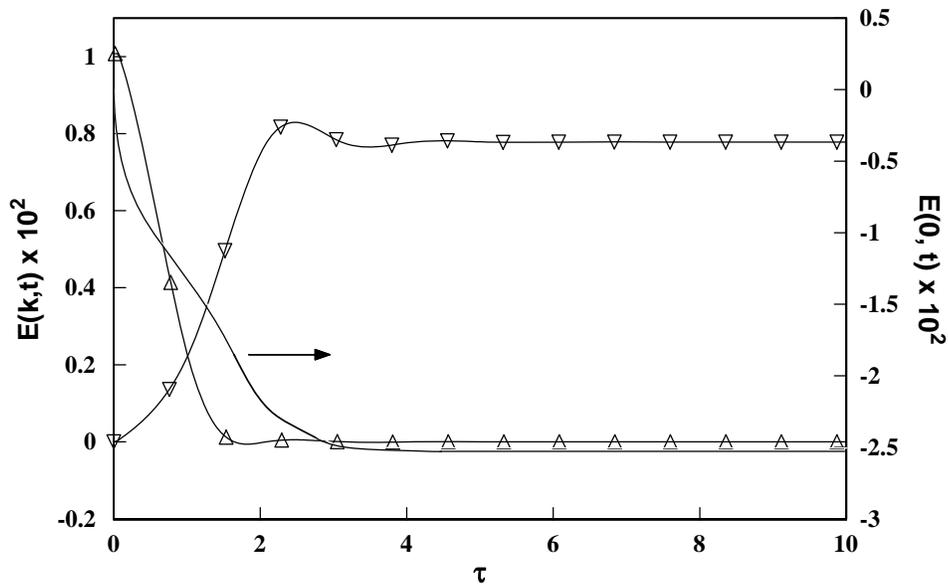

Figure 2. Evolution of the kinetic energy for Re=88.1 and h=0.5 for a single initial mode k =1.75.

—△— k = 1.75   —▽— k = 3.5   —— k = 0

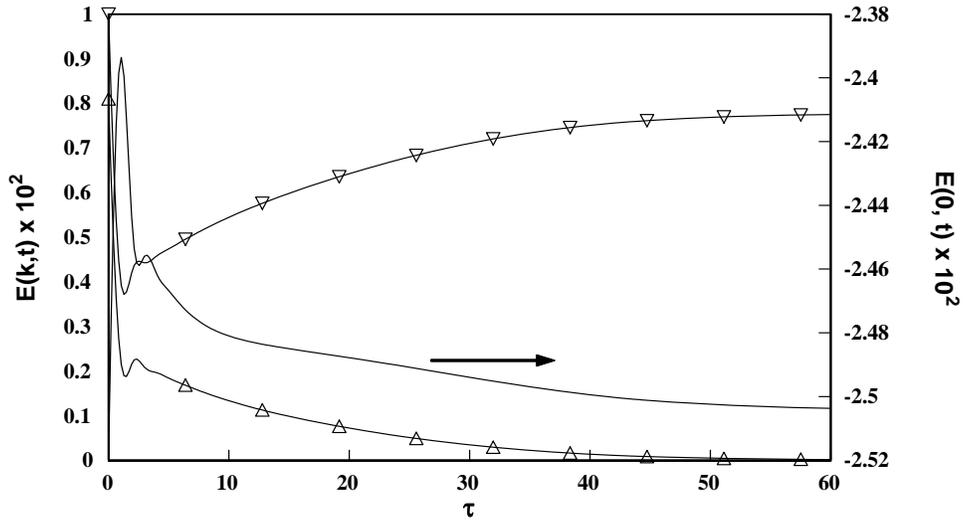

Figure 3. Evolution of the kinetic energy for Re=88.1 and h=0.5 for two initial modes at k=3.25 and 3.5. The initial amplitude at k=3.5 is larger.

△— k = 3.25    ▽— k = 3.5    ——— k = 0

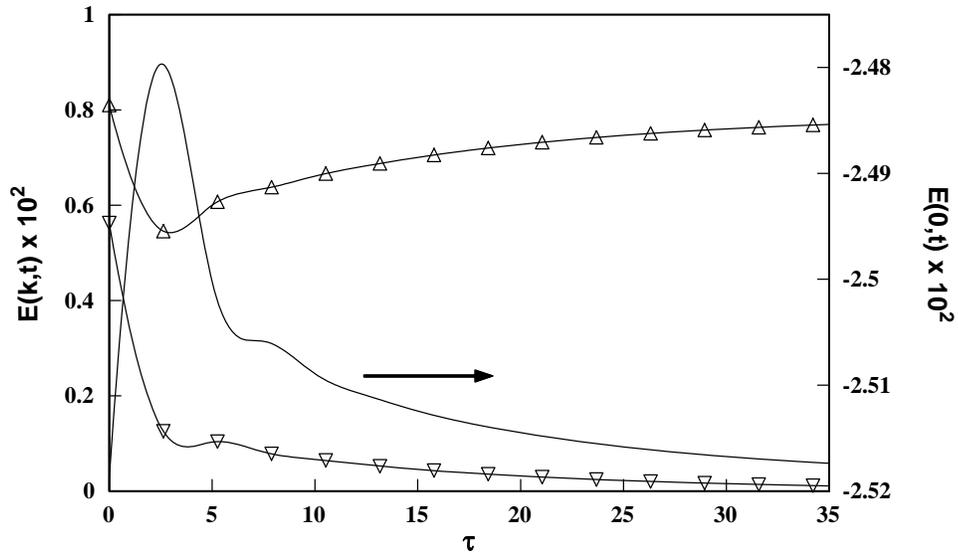

Figure 4. Evolution of the kinetic energy for Re=88.1 and h=0.5 for two initial modes at k=3.25 and k=3.5. The initial amplitude of k=3.25 is larger.

△— k = 3.25    ▽— k = 3.5    ——— k = 0

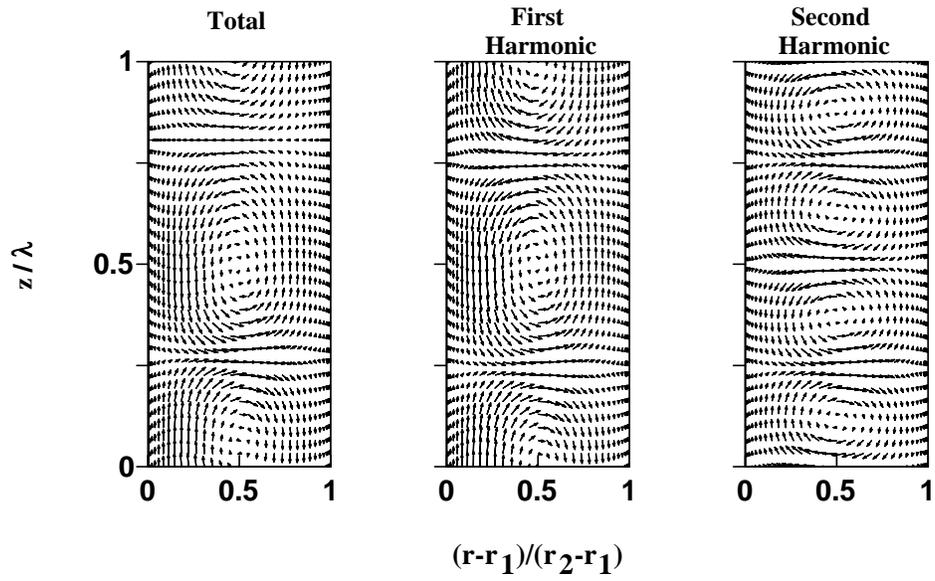

Figure 5. Projection of the velocity vectors for Taylor-vortex flow onto the r-z plane for Re=88.1, h=0.5 and k=3.5.